\documentstyle[prb,aps,multicol,psfig]{revtex}

\def\ket#1{\left| #1 \right\rangle}
\def\bra#1{\left\langle #1\right|}

\begin{document}
\draft
\title{Critical Behavior of the Random Potts Chain}
\author{E. Carlon \and C. Chatelain
and B. Berche}
\address{Laboratoire de Physique des Mat\'eriaux,\footnote{Unit\'e
Mixte de Recherche CNRS No~7556} Universit\'e Henri Poincar\'e Nancy 1, \\
B.P.~239, F - 54506 Vand{\oe}uvre les Nancy Cedex, France}

\date{\today}

\maketitle

\begin{abstract}
We study the critical behavior of the random $q$-state Potts quantum
chain by density matrix renormalization techniques. Critical exponents 
are calculated by scaling analysis of finite lattice data of short 
chains ($L \leq 16$) averaging over all possible realizations of
disorder configurations chosen according to a binary distribution. Our numerical
results show that the critical properties of the model are
independent of $q$ in agreement with a renormalization
group analysis of Senthil and Majumdar (\prl {\bf 76}, 3001 (1996)). 
We show how an accurate analysis of moments of the distribution of 
magnetizations allows a precise determination of critical exponents, 
circumventing some problems related to binary disorder.
Multiscaling properties of the model and dynamical correlation functions
are also investigated.
\end{abstract}

\pacs{05.20.-y, 05.50.+q, 64.60.Fr}

\begin{multicols}{2} \narrowtext
\section{Introduction}
\label{sec:intr}

The study of the critical properties of systems subject to quenched
randomness is of great interest because of its relevance  for many
experimental systems. This is a challenging research area which has been quite
active in the last decades. One has witnessed a growing interest in
understanding the effect of randomness (bond and site disorder), since 
Harris showed that disorder is a relevant perturbation in a system
undergoing a 
second
order phase  transition when the exponent $\alpha$
of the specific heat in the pure system is positive.\cite{harris74} 
In this context the
two dimensional Ising
model has 
attracted
considerable interest, since the Onsager
singularity of the specific heat ($\alpha=0$) makes the Harris criterion
inconclusive (for a review, see e.g. Ref.~\onlinecite{dotsenko95}).
Renormalization group studies,\cite{shalaev94} supported by Monte 
Carlo simulations,\cite{selkeshchurtalapov94} established that 
homogeneous disorder is marginally irrelevant, 
i.e. it does not modify the critical behavior of the system, except for
the appearance of logarithmic corrections.

In accordance with the Harris criterion, randomness is on the other hand 
a relevant perturbation for the $q=3$ and $q=4$ Potts models. Furthermore, 
an infinitesimal amount of disorder was shown to turn the
first
order phase transition of the $q>4$ Potts model into a 
second
order one.\cite{imrywortis79,huiberker89,aizenmanwehr89} Numerical
estimates of the critical exponents for an homogeneous disorder indeed 
showed a monotonic increase of the magnetization exponent $\beta$ with 
the value $q$ while the thermal exponent $\nu$  remains constant 
within numerical 
accuracy.\cite{cardyjacobsen97,jacobsencardy98,picco98,chatelainberche98}
Thus for the two dimensional classical Potts model, {\it homogeneous}
disorder gives rise to several different universality classes 
dependent on the value of $q$.

Other systems with quenched randomness which have been investigated in the
past years are quantum spin chains. The interest in these models has grown
considerably after a remarkable paper of Fisher~\cite{fisher92} who
derived many of the critical properties of 
the random transverse Ising chain (RTIC)
using a real space renormalization-group scheme which is claimed to be
asymptotically exact. His predictions were checked numerically by several
authors, using the mapping of the 
RTIC
onto a free fermion
model,\cite{youngrieger96} density-matrix renormalization
group~\cite{juozapavicius97} (DMRG) or quantum Monte Carlo.\cite{rieger98} 
Critical exponents calculated numerically were found in good agreement
with Fisher's results.
The renormalization scheme of Fisher was  also applied to
the $q$-state disordered Potts chain by Senthil and 
Majumdar;\cite{senthil96} they found that the critical behavior does not
depend on $q$, therefore all critical exponents should be identical to 
those of the RTIC obtained by Fisher and corresponding to the case $q=2$.
We recall that the disordered quantum Potts chain is equivalent to a
two dimensional classical Potts model with {\it correlated} disorder
(analogous to the McCoy-Wu model~\cite{mccoywu68}  corresponding
to $q=2$), where the quantum formulation is obtained from the classical 
one in the extreme anisotropic limit.\cite{kogut79}
The conclusion of a unique universality class is thus different from 
what found in the classical Potts model with homogeneous disorder 
discussed above.

In this paper, we check the predictions of a unique universality class,
independent of the number of states $q$, by studying  bulk and surface
magnetization properties. We consider both the regimes $q\le 4$ and
$q>4$, where the pure model exhibits a second and first-order transition, 
respectively. We also report an investigation of 
the multiscaling behavior of the magnetization, which strongly supports 
the asymptotic expression for the probability distribution found by Fisher 
in the 
RTIC
case.
The dynamical behavior is considered through the decay of the spin-spin
correlation function 
along
the time direction. 
All numerical results are obtained by density matrix renormalization 
group techniques,\cite{white92,book} applied to the two dimensional classical 
Potts model with correlated disorder. 
We use Nishino's~\cite{nishino95} version of White's DMRG 
algorithm,\cite {white92} where one renormalizes classical transfer matrices. 
The exponents are calculated from finite size scaling 
analysis of finite lattice data for chains of lengths $L \leq 16$ 
with exact enumeration over all possible ($2^L$) disorder realizations,
obtained from a binary distribution. The justification for this type 
of approach will be presented in more detail in the text.

Our numerical results are fully consistent with the conclusion of Senthil 
and Majumdar,\cite{senthil96} i.e. the $q$-state Potts model has critical 
properties which do not depend on $q$. Moreover, dynamical properties which 
are not accessible by Fisher's renormalization group, do not seem to depend 
on $q$.

The paper is organized as follows: In Sec.~\ref{sec:mod}, the model is
presented and we give a justification for the choice of the numerical methods
that are used.  Section~\ref{sec:exp} is devoted to the estimates of the
critical exponents. Multi-fractality is investigated in Sec.~\ref{sec:multi}.
The dynamical spin-spin correlation functions are studied in
Sec.~\ref{sec:corr}. Finally, a brief summary and an outlook will be
presented in Sec.~\ref{sec:con}.

\begin{figure}[h]
\centerline{\psfig{file=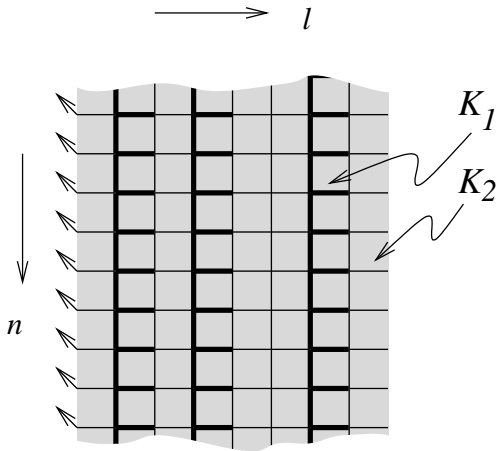,height=6.5cm}}
\vskip 0.2truecm
\caption{Layered structure of the model
with coupling constants $K_1$ and $K_2$ taken from
the distribution (\ref{eq2}) and indicated by thick
and thin bonds, respectively.
The arrows on the left 
indicate fixed boundary conditions, while spins on the right surface 
are left free.}
\label{fig5}\end{figure}

\section{Model and numerical methods}
\label{sec:mod}

We consider a randomly layered (see Fig.~\ref{fig5}) two-dimensional 
classical $q$-state 
Potts model on a square lattice. The model is defined by the Hamiltonian
	\begin{equation}
	-\beta{\cal H} = \sum_{l=1}^L K_l \sum_n
        \left [\delta_{\sigma_{l,n},\sigma_{l+1,n}}
	+\delta_{\sigma_{l,n},\sigma_{l,n+1}}\right],
	\label{eq1}\end{equation}
where the spins can take the values $\sigma_{l,n}=0,\ldots, q-1$ and 
the exchange couplings $K_l$ are randomly chosen according to 
the binary distribution
	\begin{equation}
	{\cal P}(K_l) = {1\over 2}\left[\delta(K_l-K_1)+\delta(K_l-K_2)\right].
	\label{eq2}\end{equation}
Since the model is equivalent to the random quantum Potts chain studied by 
Senthil and Majumdar,\cite{senthil96} we use the quantum chain's vocabulary.
In the case $q=2$, the system reduces to the 
RTIC. 

The particular arrangement of coupling constants, i.e. the fact
that vertical couplings
are identical to adjacent horizontal couplings to their right side (see
Fig. \ref{fig5}), and  that 
both couplings $K_1$ and $K_2$
have equal probabilities 
(see
Eq.~(\ref{eq2})), 
makes the model self-dual. The location of the critical point is 
known exactly:\cite{wisemandomany95}
\begin{equation}
\left(e^{K_1}-1\right)\left(e^{K_2}-1\right)=q.
\label{eq3}\end{equation}
The strength of the disorder is monitored by the ratio
$r=K_1/K_2$, $r=1$ corresponding to the pure model.

Although the DMRG method is capable of dealing with very large systems
(of few hundreds of sites) with remarkable accuracy, some care has
to be taken when working with disordered systems. 
The infinite lattice version of the algorithm is known to give incorrect
results,\cite{prokofev98}
especially at strong disorder; it is {\it essential}~\cite{rapsch} when dealing with 
inhomogeneous systems to use the finite lattice DMRG algorithm~\cite{white92} which
is designed to determine accurately properties of finite systems (In 
our calculations we typically used three full
sweeps through the lattice).
It is furthermore well known that due to the lack of
self-averaging,\cite{aharonyharris96,wisemandomany98a}
an
insufficient
number
of samples 
yields
a wrong estimation of the average over randomness and might lead to 
typical values instead of the expected average quantities.\cite{derrida84}
As we aim to an accurate determination of critical exponents, we have 
chosen to restrict ourselves to relatively narrow strips ($L \leq 16$ which
turns out to be large enough, since corrections to scaling will appear
very small)
for which we were able to do an exact enumeration over all disorder
realizations. As the system size is small compared to typical DMRG
calculations a limited number of states kept ($m \leq 16$) is sufficient
to obtain accurate results.

The setup of the calculation follows 
Ref.~\onlinecite{carlonigloi98}, where the DMRG method was used
to calculate exponents and magnetization profiles for the pure Potts model.
The $Z_q$-symmetry of the Potts model has been broken by introducing 
fixed--free boundary conditions:
We fixed the spin at one edge of the system to a value $\sigma_{0,n} = 0$, 
while at the
other edge the spin is free. This type of boundary conditions induces a
finite magnetization in the system~\cite{rq}  from which surface and bulk critical 
exponents can be calculated using finite size scaling analysis.

\begin{figure}[h]
\centerline{\psfig{file=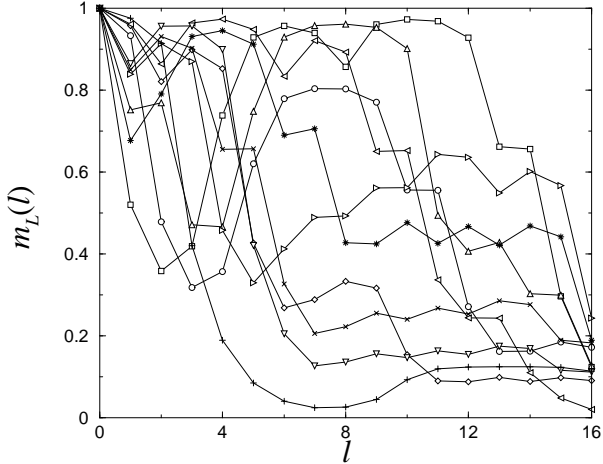,height=6.5cm}}
\vskip 0.2truecm
\caption{Magnetization profiles for ten random configurations of exchange
couplings at $q=3$, $r=2$, $L=16$. Error bars are smaller than the symbols.}
\label{fig4}\end{figure}

Figure~\ref{fig4} shows magnetization profiles (defined by Eq.~(\ref{eq4}))
of $10$ samples for the
three-state Potts model with $L = 16$ and with disorder amplitude $r=2$.
Estimated error bars are much smaller than symbol sizes; averaging over all disorder
realizations produces the smooth profiles shown in Fig. \ref{fig1}.
In this procedure the only source of error comes from the truncation on 
the DMRG basis, this error is quite small even for $m=16$ states kept 
as we restricted 
ourselves to small systems.

\section{Critical exponents}
\label{sec:exp}

\subsection{Finte-Size scaling}
\label{sec:ffs}

We used the standard order parameter of the Potts model:
	\begin{equation}
	\left[ m_L(l)\right]_{\rm av}
	= {q \left[ \langle\delta_{\sigma_{l,n},0}\rangle \right]_{\rm av}
        - 1\over q - 1},
	\label{eq4}\end{equation}
where $\langle \dots\rangle $ denotes the thermal average and 
$\left[\dots\right]_{\rm av}$ the average over randomness.
Standard finite size techniques have been used to estimate 
the bulk and surface critical exponents
	\begin{equation}
	[  m_L^b ]_{\rm av}\sim L^{-x_b},		\quad
	[  m_L^s ]_{\rm av}\sim  L^{- x_1},
	\label{eq5}\end{equation}
where we have defined $[  m_L^b ]_{\rm av} \equiv
[  m_L(L/2) ]_{\rm av}$ and $[  m_L^s ]_{\rm av} 
\equiv [  m_L(L) ]_{\rm av}$, which are the average magnetizations
in the middle of the strip and at the free edge.
{From} finite size data we calculated the approximants, defined
as
	\begin{equation}
	x_b(L) = -{\ln [  m_{L+1}^b ]_{\rm av}
		- \ln [  m_{L-1}^b ]_{\rm av}
		\over \ln (L+1)-\ln (L-1)}
	\label{eq6}\end{equation}
 in the bulk case  and similarly  for the surface approximants.
In the limit $L \to \infty$ the approximants scale towards the bulk
and surface exponents $x_b$ and $x_1$.

Table~\ref{table1} reports the finite size approximants for the surface
exponents for various values of $q$ and of disorder ratios $r$. 
The extrapolation towards the thermodynamic limit was done using the
BST algorithm.\cite{henkelschutz88}
The boundary magnetization exponent, $x_1$, exhibits a striking convergence
for all values of $q$ and all disorder amplitudes, but for crossover
effects discussed below. It is clearly independent 
of 
$q$, and in good agreement with the RTIC value $x_1 = 1/2$.
The best numerical results are obtained for $q=2$, as the truncation error
in the DMRG method is the smallest for this case, furthermore the surface
exponent for the pure and disordered cases coincide for $q=2$ (they are both
equal to $1/2$), therefore no crossover effects are to be expected. 

\end{multicols} \widetext

\vbox{
\begin{table}
\caption{Approximants of the surface exponent  for various $q$ and $r$.
The last row shows the values extrapolated with the BST algorithm. We note that
at $q=6$ the disorder amplitude being quite small, crossover effects from the
pure model value $x_1=3$ slow down the convergence. 
}
\vskip 0.2truecm
\begin{tabular}{r|c|c|c|c|c|c|c|c
}
 & \multicolumn{2}{c|}{$q = 2$}
 & \multicolumn{3}{c|}{$q = 3$}
 & \multicolumn{2}{c|}{$q = 4$}
 & $q = 6$
\\
\hline
$L$ & $r =2$ & $r = 8$ & $r = 2$ & $r = 4$ & 
$r = 8$ & $r = 5$ & $r = 10$ & $r = 1.5$ 
\\
\hline
 5  & 0.41608413 & 0.44950692 & 0.46559328 & 0.46393358 & 0.46844808 &
      0.47671724 & 0.47741488 & 0.55719868 
\\
 7  & 0.43773154 & 0.46307866 & 0.48454642 & 0.47694537 & 0.47932515 &
      0.48718127 & 0.48819610 & 0.56816316 
\\
 9  & 0.45043349 & 0.47083067 & 0.49365387 & 0.48353868 & 0.48475158 & 
      0.49161650 & 0.49204019 & 0.56481411 
\\
11  & 0.45880966 & 0.47587114 & 0.49851791 & 0.48742293 & 0.48800690 &
      0.49383119 & 0.49387188 & 0.56108433 
\\
13  & 0.46475540 & 0.47941816 & 0.50127872 & 0.48993505 & 0.49014411 &
      0.49481028 & 0.49435771 & 0.54974464 
\\
15  & 0.46919708 & 0.48205234 &      -     & 0.49166836 & 0.49150054 &
      - &      -     &      -     
\\
\hline
$\infty$ 
    & 0.50000(1) & 0.5000(1)  & 0.498(4)   & 0.499(1)    & 0.500(1)  &
      0.496(10)  & 0.50(5)    &    -
\end{tabular}
\label{table1}
\end{table}
}

\begin{multicols}{2}\narrowtext

We recall that the values of the surface exponents for the pure Potts model
are $x_1 = 2/3$ for $q=3$, $x_1 = 1$ for $q=4$ and $x_1 = 3$ for $q > 4$.
The last exponent has also been recently calculated by DMRG 
methods~\cite{igloicarlon99} (the surface transition is continuous 
even if the bulk has a first order transition). In the case $q=6$, 
$r=1.5$, one notices
that the approximants approach the value $1/2$ in a non-monotonic 
fashion (the same situation occurs with $q=4$, $r=2$). This case corresponds to
the weakest disorder considered and the non-monotonicity is most 
likely due to crossover effects
from the exponent of the pure system ($x_1=3$). It is not possible 
to extrapolate
the data in the last column of Table~\ref{table1} with the BST method, since
 the largest
size analyzed is still in the crossover region. However we can give an estimate
of the exponent using the analysis of the moments presented in 
Section IV B; from such an analysis we find 0.49(3) in good agreement with the
value of the RTIC, $1/2$. We also mention that calculations have 
been performed at $q=6$, $r=4$. The approximants then do not suffer from so
strong crossover effects, an observation which supports
the explanation given above.

Table~\ref{table2} shows the approximants of the bulk exponent
obtained from Eq.~(\ref{eq6}) for various $q$ and $r$. As it can
be seen the approximants show a non-monotonic
behavior, with oscillations which are stronger as the disorder
ratio increases.
These oscillations, which were already observed in the 
RTIC,\cite{igloirieger97}
are due to the choice of a binary distribution of exchange 
couplings
which introduces an energy scale $K_1 - K_2$ in the problem; in other
types of distributions, e.g. continuous ones, these oscillations are
not present.\cite{igloirieger97}

Lattice extrapolation techniques, as the BST method, are not able to
deal with oscillating corrections to scaling. 
We recall that the BST algorithm generates from an original series 
of $N$ approximants new series of $N-1$, $N-2$ \ldots approximants 
at each iteration which are expected to have faster convergence towards 
the asymptotic value. In the BST analysis we found that even when starting
from a monotonic series of data at weak disorder ratios, as in the cases 
$q=2$, $r=2$, the next series generated by the algorithm do not converge
at all, but they show oscillating behavior. Therefore the BST method 
turns out to be useless in the analysis of the bulk exponent data.
A better way to get a precise estimate of bulk exponent is based on the
analysis of moments, which will be explained in Section \ref{sec:manal}.

In any case the approximants of Table~\ref{table2} for different values
of $q$ are rather close to each other, which supports again the idea
of $q$-independence of their values. We recall that for the pure Potts 
model the bulk exponent is $x_b=1/8$ for $q=2,4$ and $x_b=2/15$
for $q=3$. For $q > 4$ the bulk transition is first order.

Fisher's prediction for the bulk 
exponent is:\cite{fisher92}
	\begin{equation}
	x_b = {3-\sqrt 5\over 4}\approx 0.19098.	\quad
\label{eq7}\end{equation}

\begin{figure}[h]
\centerline{\psfig{file=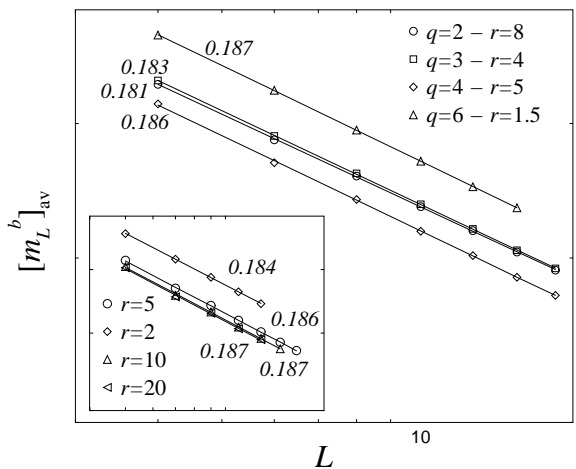,height=6.5cm}}
\vskip 0.2truecm
\caption{Power-law behavior of the bulk magnetization with the lattice size
for several values of $q$ and $r$ (inset shows different disorder amplitudes
for $q=4$). The values indicated correspond to a simple power-law fit.}
\label{fig6}\end{figure}

Figure~\ref{fig6} shows a log-log plot of bulk magnetization vs. 
system size. The data correspond to different $q$ and $r$ and 
are all roughly parallel with slopes in the range $0.183 - 0.187$
indicated in the figure and obtained by a simple linear fit. In 
the scale of the figure oscillations are not visible and the
estimated slopes correspond  to the average of the
approximants of Table~\ref{table2}.

\end{multicols} \widetext
	
\vbox{
\begin{table}
\caption{Approximants of the bulk exponent  for various $q$ and $r$.}
\vskip 0.2truecm
\begin{tabular}{r|c|c|c|c|c|c|c|c
}
 & \multicolumn{2}{c|}{$q = 2$}
 & \multicolumn{3}{c|}{$q = 3$}
 & \multicolumn{2}{c|}{$q = 4$}
 & $q = 6$
\\
\hline
$L$ & $r =2$ & $r = 8$ & $r = 2$ & $r = 4$ & $r = 8$ & 
$r = 5$ & $r = 10$ &
$r = 1.5$ 
\\
\hline
 5  & 0.15352463 & 0.18540035 & 0.17119099 & 0.18598351 & 0.19697064 &
      0.19644591 & 0.20327857 & 0.18415244 
\\
 7  & 0.15902213 & 0.17126232 & 0.17606416 & 0.17698031 & 0.17098093 &
      0.17386925 & 0.16655933 & 0.18858150 
\\
 9  & 0.16285921 & 0.18653590 & 0.17971019 & 0.18728278 & 0.19301242 &
      0.19257056 & 0.19639122 & 0.18942360 
\\
11  & 0.16572102 & 0.17921449 & 0.18137938 & 0.18203550 & 0.19536097 &
      0.18054979 & 0.17782952 & 0.18957220 
\\
13  & 0.16802871 & 0.18804307 & 0.18299470 & 0.18816428 & 0.18974244 &
      0.19214731 & 0.19569786 & 0.18635074 
\\
15  & 0.16991419 & 0.18254668 &      -     & 0.18411149 & 0.19515526 & 
      0.18319909 &      -     &      -     
\\
\end{tabular}
\label{table2}
\end{table}
}

\begin{multicols}{2}\narrowtext

We mention that, surprisingly, we observed a weak dependence of the 
effective critical exponents with the amplitude of disorder $r$ in 
the range $[2;20]$, while important cross-over effects with the 
strength of the disorder were observed, e.g. by Picco,\cite{picco98}
in the case of a two-dimensional Potts model with a homogeneous disorder. This is an evidence of the strongly 
attractive character of the random fixed point.              

\subsection{Transverse magnetization profiles}
\label{sec:tmp}

Conformal invariance techniques are extremely accurate for studying second 
order phase transitions in isotropic pure systems. 
Conformal symmetry
requires translation, rotation and scale invariance; obviously a single disorder
realization does not have such properties, but it is plausible that conformal symmetry
is restored after averaging 
 over different disorder
 configurations. Indeed in recent investigations, conformal 
 mappings of correlation functions and order 
parameter profiles
 have been successfully applied in two-dimensional systems with homogeneous 
 disorder.\cite{chatelainberche98b}
The case of correlated disorder, studied in this paper, is quite different
as the system, even after disorder average, exhibits 
infinite anisotropy at the random fixed point (see Sec.~\ref{sec:corr}). 
Therefore conformal invariance should not hold.
In spite of these restrictions, the magnetization profiles of the 
RTIC
with fixed-free boundary conditions have been quite
satisfactorily fitted with the conformal expression,\cite{igloirieger97},
obtained e.g. by Burkhardt and Xue \cite{burkhardtxue91}
	\begin{equation}
	[  m_L(l) ]_{\rm av}={\cal A}(\pi/L)^{x_b}
        [\sin\pi \zeta]^{-x_b} [\cos\pi \zeta/2]^{x_1},
	\label{eq10}\end{equation}
where $\zeta=(l-1/2)/L$.
The success of such an approach  probably results from the geometrical 
constraints
induced by the strip shape of the system combined to scaling arguments:
In this sense, if Eq.~(\ref{eq10}) turns out to be valid also for the 
disordered Potts chain, one should conclude that such a form 
of profile is not a strict
consequence of conformal invariance, which  does not hold in anisotropic 
critical systems,\cite{henkel92}
but  follows from more general grounds.

\begin{figure}[h]
\centerline{\psfig{file=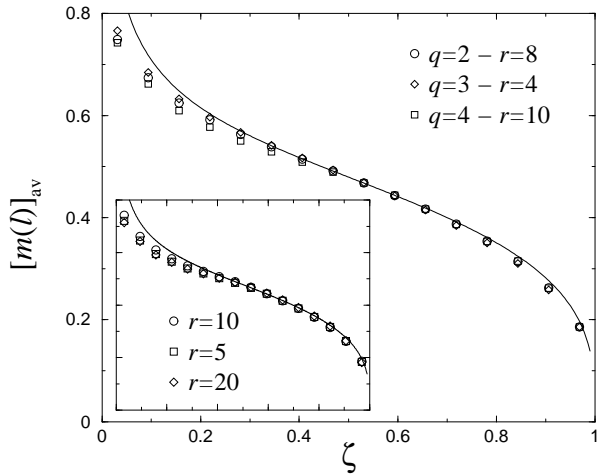,height=6.5cm}}
\vskip 0.2truecm
\caption{Magnetization profiles for several values of $q$ and $r$ 
(inset for $q=4$ and several disorder amplitudes)
	and the conformal expression with the Fisher exponents (solid line)
	with only the amplitude as a free parameter.}
        \label{fig1}\end{figure}

In this Section, we indeed ask the question whether Eq.~(\ref{eq10}) correctly
describes the transverse magnetization profiles in the
random Potts chain. Senthil and Majumdar~\cite{senthil96} indeed argued 
that the scaling functions should be  independent of $q$. The quantity
$m(l) L^{x_b}$ being such a scaling function, if Eq.~(\ref{eq10})
holds in the 
RTIC 
chain when $L\to\infty$, it should also be valid in the same limit in
 the Potts case.
Unfortunately, in our case, the largest strip being $L=16$, 
strong lattice effects make the
fit to expression~(\ref{eq10}) impossible in practice (Fig.~\ref{fig1}). 
The same observation was reported for the 
RTIC,
but these finite-size effects vanish for the largest systems that can be 
reached using the free fermion mapping in the Ising case.
If Eq.~(\ref{eq10}) is valid, the quantity 
	\begin{equation}
	x_1(L)={\ln [  m_L(3L/4) ]_{\rm av}-
	\ln [  m_L(L/4) ]_{\rm av}\over \ln(\sqrt 2-1)}
	\label{eq11}\end{equation}
	should converge towards the surface exponent $1/2$.
The results are shown in Table~\ref{table3}.
This method cannot be considered as an exact determination
of $x_1$ but it provides an indication of the possible 
validity of~(\ref{eq10}) for this system.
The bulk exponent $x_b$ also follows from a three-point 
expression analogous to
Eq.~(\ref{eq11}) and should be extracted
from the magnetization profile, but our data are 
not accurate enough to allow this calculation.

\vbox{
\begin{table}
\caption{Approximants of the surface exponent for various $q$ and $r$
calculated with (\ref{eq11}). The sizes in bold face correspond to chain 
for which there is a lattice point $l$ such that $(l-1/2)/L=1/4$, while the
other data were obtained after fitting the finite-size profile.
 The results are compatible with the expected
value of 0.5, seemingly supporting the expression of the transverse profile.}
\vskip 0.2truecm
\begin{tabular}{r|c|c|c|c
}
$L$ & $q=2 - r=8$  & $q=3 - r=4$ &  
$ q=4 - r=5$ & $ q=6 - r=4 $
\\
\hline
4 & 0.5161 & 0.5195 & 0.5505 & 0.5709 \\
 \bf 6 & 0.5094 & 0.5237 & 0.5494  & 0.5711
\\
8 & 0.5039 & 0.5224 & 0.5411 & 0.5607 \\
\bf 10 & 0.4996 & 0.5210 & 0.5341 & 0.5511
\\
12 & 0.4995 & 0.5202 & 0.5314 & 0.5422 \\
\bf 14 & 0.5010 & 0.5206 & 0.5310 & 0.5262\\ 
16 & 0.5000 & 0.5187 & 0.5269 & - \\
\end{tabular}
\label{table3}
\end{table}
}

\section{Multi-fractality}
\label{sec:multi}

\subsection{Probability distribution}
\label{sec:pdist}

According to Fisher, the probability distribution of the 
surface magnetization $m_L^s$ for a strip of
width $L$ is expected to behave asymptotically as:

	\begin{equation}
	{\cal P}(  m_L^s )
	\sim {1\over\sqrt L}\tilde p\left({\ln   m_L^s \over\sqrt L}\right).
	\label{eq12}\end{equation}

\begin{figure}
\centerline{\psfig{file=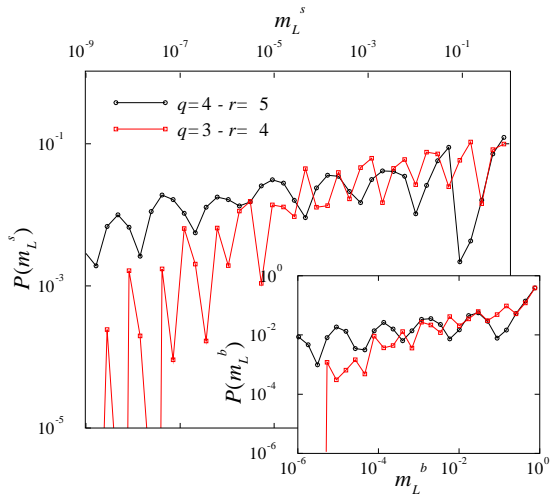,height=6.5cm}}
\vskip 0.2truecm
        \caption{Probability distribution of the surface and bulk (inset) 
        magnetizations
	defined with 64 bins ($q=3$ (squares) and $q=4$ (circles), $L=16$).}
        \label{fig2}\end{figure}

This expression was successfully used for the 
RTIC
by Igl\'oi and
Rieger to rescale the probability distribution of the surface
magnetization for several strips of different widths $L$ in 
Ref.~\onlinecite{igloirieger97}. The scaling function $\tilde p$ 
was found to have a power-law decay in the
RTIC.
Forgetting about the oscillations, the average
behavior in Fig.~\ref{fig2} is compatible with this observation.

The results of Senthil and Majumdar~\cite{senthil96} state that the 
scaling function $\tilde p$ is independent of $q$ in the thermodynamic limit. Here again, the 
widths of our strips are too small to give a smooth probability 
distribution, as shown in Fig.~\ref{fig2}. The Figure shows that for the
events  which determine the average critical
properties, i.e. those corresponding to the dominant contributions
to the magnetization, the probability distribution is weakly dependent on $q$,
in accordance with universality of Eq.~(\ref{eq12}).
On the other hand, for the rare samples for which $m_L^s$ or
$m_L^b\le 10^{-5}$, 
Eq~(\ref{eq12}) breaks down.
 It is however possible to 
test expression~(\ref{eq12}) by analyzing the moments of the 
distribution. 

\subsection{Moments analysis}
\label{sec:manal}

In this section we analyze moments of bulk and surface magnetization 
distributions. This analysis will help to avoid the problems related to the oscillations of the finite 
size estimates of the bulk exponent which, as shown in the previous 
section, hindered a reliable extrapolation of such quantity. Moments 
analysis was recently used  in studies of 
self-organized critical 
systems.\cite{demenechstellatebaldi98,demenechstellatebaldi99} It 
was shown that such method provides an accurate means of investigation 
of subtle aspects that seem to 
emerge for such models.

\begin{figure}
\centerline{\psfig{file=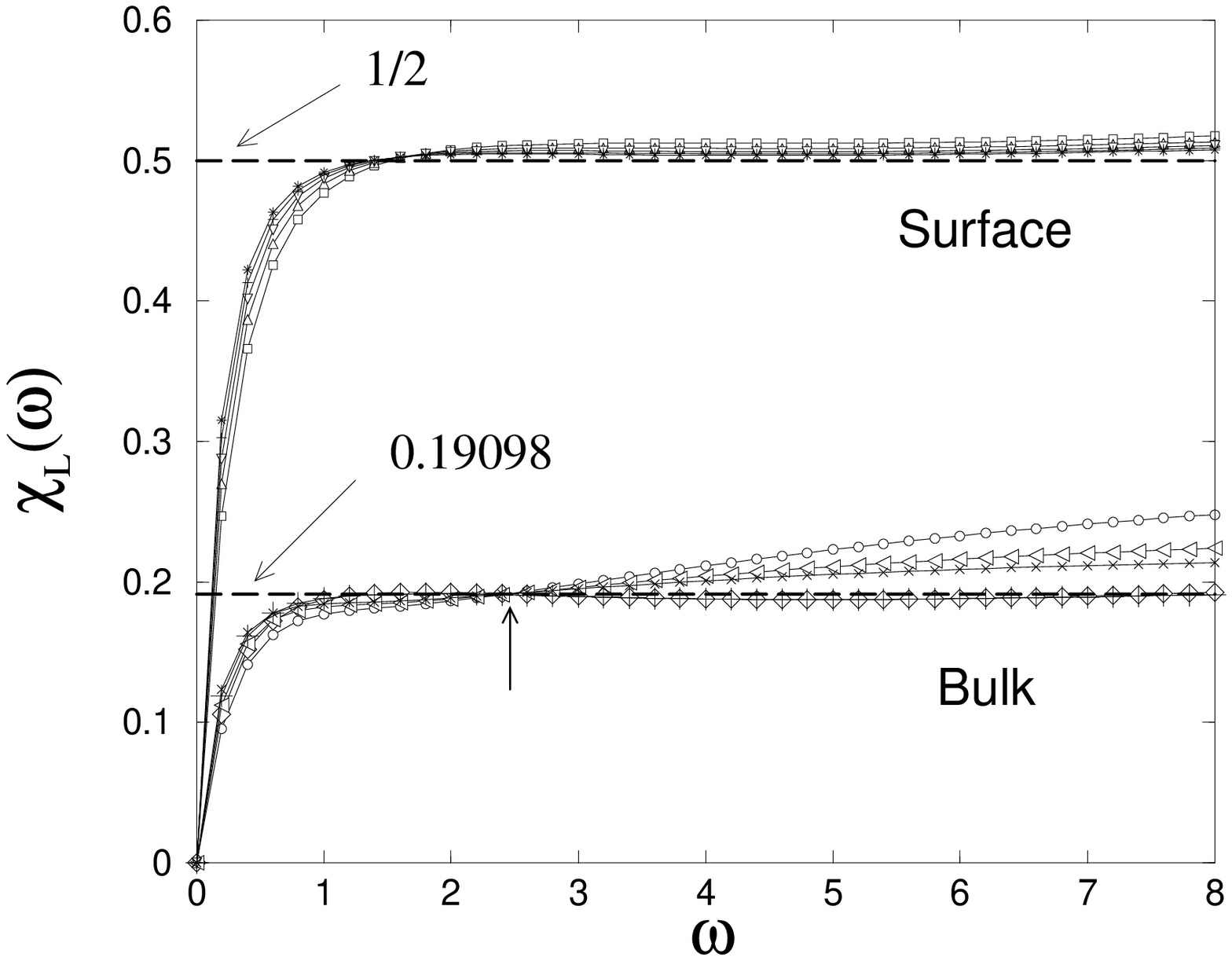,height=6.5cm}}
\vskip 0.2truecm
        \caption{Approximants of the bulk and surface magnetization 
        critical exponents vs the order of the moments used to 
        calculate them. The different symbols correspond to different 
        sizes from 5 to 15 ($q=3$, $r=4$).}
        \label{fig7}\end{figure}

In our case we are interested in the scaling of the asymptotic behavior 
of the moments. For instance, in the case of the surface or bulk 
magnetization we expect:
	\begin{equation}
	{[{  (m_L)^\omega }}]_{\rm av} \sim 
        L^{-\chi (\omega )}.
	\end{equation}
The exponent $\chi (\omega )$ generalizes the definitions of 
Eq.(\ref{eq6}), which correspond to $\omega = 1$.
If the system is self-averaging, we expect a linear dependence of $\chi 
(\omega )$ with respect to $\omega$.\cite{rq2} 
At a fixed size $L$, we consider 
the finite-size approximants of $\chi (\omega )$ defined as usual by:
	\begin{equation}
	\chi_L (\omega ) = - \frac{\ln {[ {  (m_{L+1}   )
        ^\omega }}]_{\rm av} - \ln {[ { ( m_{L-1}    )
        ^\omega }]_{\rm av}}} {\ln ( L + 1) - \ln ( L - 1 )}.
	\end{equation}
Plots of both quantities for $L = 5, 7, \ldots 15$ 
and $q=3$, $r = 4$
are shown in Fig.~\ref{fig7}. 
The values at $\omega = 1$ are those shown in the 
tables~\ref{table1} and \ref{table2}. 
The fact that $\chi_L (0) = 0$ is a consequence of
the normalization of the probability distribution.

We notice that in both the surface and bulk case $\chi_L (\omega )$ 
for large $L$ seems to be a constant independent of $\omega$; this 
constant equals the surface or bulk exponent. 
This can be understood if one assumes the 
probability distribution given
in Eq.~(\ref{eq12}).
As a consequence of this one indeed obtains:
	\begin{eqnarray}
	{[ {  (m_L^s)^\omega }]_{\rm av}} = 
        \frac{1}{\sqrt{L}} \,\, 
	\int_0^1 d m_L^s \,\, (m_L^s)^\omega \,\, \tilde{p} 
        \left( \frac{ \ln m_L^s }{\sqrt L} \right) \nonumber \\
	= \frac{1}{( 1 + \omega ) \sqrt{L}} 
	\int_0^1 d x \,\, \tilde{p} \left( \frac{ \ln x} 
        {( 1 + \omega ) \sqrt{L}} \right),
	\label{scalgamma}
	\end{eqnarray}
where we have used the change of variables $x = (m_L^s )^{\omega + 1}$.
We notice from this formula that $\omega$ does not modify the 
exponent of $L$, but enters only as a prefactor of $\sqrt{L}$ and consequently
all the moments should have the same scaling dimension $\chi(\omega)$.
This is what observed in Fig.~\ref{fig7} where
 the calculated $\chi_L (\omega )$ does not seem to 
depend on $\omega$, therefore our numerical results are consistent with 
a probability distribution of the form of Eq.~(\ref{eq12}). 
It is also important to stress that results would have been very 
different for a probability distribution with another scaling variable, 
e.g. $m_L^s/ \sqrt{L}$.
We note that in Eq.~(\ref{scalgamma}), the last integral seems to be nothing 
but a normalization, but as we mentioned in the previous Section
the expression
used for the probability distribution is only asymptotically valid 
for the dominant
events, and thus the true normalization of the distribution differs from  
Eq.~(\ref{scalgamma}).

An 
interesting
consequence of these observations is that one could 
calculate surface or bulk exponents from an arbitrary value of 
$\omega$. From the point of view of the finite size scaling analysis 
we consider values of $\omega$ where the dependence on the system size 
$L$ is the smallest. 

For the data referring to the bulk exponent and shown in the figure 
we see that around $\omega = 2.5$ (this point is indicated by a thick 
arrow in the figure) finite size effects are very small, all 
$\chi_L (\omega )$ seem to intersect into a single value; for which 
we estimate $x_b = 0.1905(5)$ in excellent agreement with the magnetic 
exponent obtained by Fisher $(3 - \sqrt{5})/4 \approx 0.19098 \ldots$
and also indicated by a thick dashed line in the figure.

\vbox{
\begin{table}
\caption{Calculated bulk exponent from generalized scaling of the
moments.}
\vskip 0.2truecm
\begin{tabular}{c|c|c|c}
$Q = 2$, $r = 2$ & $Q = 2$, $r = 8$ & $Q = 3$, $r=4$ & $Q=6$, $r=1.5$ \\
\hline
  0.189(3)       &  0.196(3)         &    0.1910(5)   &  0.190(2)
\end{tabular}
\label{genscaling}
\end{table}
}

Table~\ref{genscaling} collects the values of bulk exponents calculated 
with this method. For all cases studied we find a value of $\omega$
for which $\chi_L (\omega )$ depends only very weakly on $L$; if the 
intersection of the $\chi_L (\omega )$ for different $L$ is sharp (as
in the example shown in Fig.~\ref{fig7}) one arrives to a very good estimate
of the exponent.

We should also stress that, as the system size enters in 
Eq.~(\ref{scalgamma}) together with $\omega$ in the form of a 
scaling variable $(1 + \omega ) \sqrt{L}$. This suggests that 
higher moments of the magnetizations should show weaker finite size
corrections, as increasing $\omega$ has the same effect of enlarging
the system size. This is only partially true as Eq.~(\ref{scalgamma})
describes only the leading scaling behavior in $1/L$, and it does not include further
corrections to scaling terms.

\section{Correlation functions}
\label{sec:corr}

In this Section, we deal with the dynamical properties 
of the random Potts chain. We thus focus here 
on correlation functions in the temporal direction, which are 
defined, in the quantum formalism, by: 
	\begin{equation}
	[G_{\sigma\sigma}(t)]_{\rm av}
	=\left[{\bra 0 \delta_{\sigma_i,0}\ e^{-{\cal H}t}
        \delta_{\sigma_i,0}\ket 0
	\over \bra 0\ e^{-{\cal H}t}\ket 0}\right]_{\rm av},
	\label{eq16}\end{equation}
where $\ket 0$ denotes the ground state.
Fisher's renormalization group does not provide any information 
about this quantity. However, Igl\'oi and Rieger by means of scaling 
arguments derived the following asymptotic behavior: 
	\begin{equation}
	[G_{\sigma\sigma}(t)]_{\rm av}\sim (\ln t)^{-2x_b}.
	\label{eq20}
	\end{equation}
This asymptotic decay of the correlation functions is unusually 
slow with respect to usual power-law behavior expected at 
criticality. This can be understood more
easily in the classical version of the model and it is due to
the fact that disorder along the transfer direction is strongly
correlated~\cite{igloirieger97} (layered disorder) and consequently the model displays
anisotropic critical behavior. It means that at the critical point, distances
measured in space and time directions are connected through power laws involving
the dynamical exponent $z$: $t\sim l^z$. The homogeneity assumption for the 
correlation functions thus becomes
\begin{equation}
	[G_{\sigma\sigma}(l,t)]_{\rm av}=b^{-2x_b}G(l/b,t/b^z),
	\label{eqGz}
\end{equation}
where $b$ is a rescaling factor. The choice $b=t^{1/z}$ thus leads to the usual 
power-law decay
\begin{equation}
	[G_{\sigma\sigma}(l,t)]_{\rm av}\sim t^{-2x_b/z}g(l/t^{1/z}),
	\label{eqGzbis}
\end{equation}
while in the case of infinite anisotropy ($z\to\infty$) one should set $b=\ln t$
and Eq.~(\ref{eq20}) follows.

In the classical formulation of the problem followed in this
paper, the dynamical correlation function is defined by:
	\begin{equation}
	[ G_{\sigma \sigma} (n) ]_{\rm av} = \left[
	\frac{ \langle v_0 | \delta_{\sigma_i, 0} 
        T^n \delta_{\sigma_i, 0} | v_0 \rangle }
	{ \lambda_0^n } \right]_{\rm av},
        \label{classicalcorrelation}
	\end{equation}
where $\lambda_0$ and $| v_0\rangle  $ are the largest eigenvalue 
and the corresponding eigenvector of the transfer matrix $T$. 
For the calculation, we used free boundary conditions on both 
edges (the magnetization defined by Eq.~(\ref{eq4}) is thus zero 
in the whole strip) and we restricted ourselves to the spin in 
the middle of the system. Since large strip widths are needed in 
order to avoid finite size effects, average are now performed 
with $\simeq 25000$ different samples.

\begin{figure}
\centerline{\psfig{file=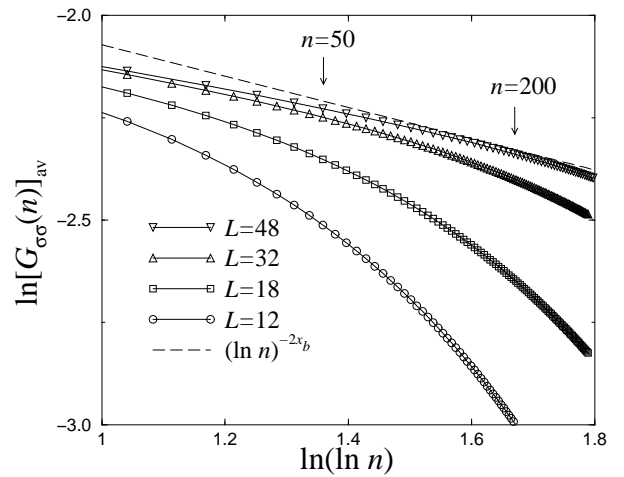,height=6.5cm}}
\vskip 0.2truecm
        \caption{Dynamical spin-spin correlation functions for the RTIC
        ($q=2$) at
	$r=4$ for several strip widths. The number of samples is 25000 for
	$L=18$ and 24000 for $L=32$ and 48.}
        \label{fig8}\end{figure}

Figure~\ref{fig8} shows the correlation function calculated by DMRG 
techniques following Eq.~(\ref{classicalcorrelation}) for $q=2$ and 
$r=4$. We plot the logarithm of the correlation function vs. the double 
logarithm of the distance along the time direction (called $n$ 
in Fig.~\ref{fig5}); the dashed
line is the asymptotic slope predicted by Eq.~(\ref{eq20}), where we
have used Fisher's exponent given in Eq.~(\ref{eq7}). This asymptotic 
behavior seems to be in agreement with our numerical data for the largest
system size analyzed ($L = 48$) for which we took $m=16$ and $24000$ disorder
realizations.

\begin{figure}
\centerline{\psfig{file=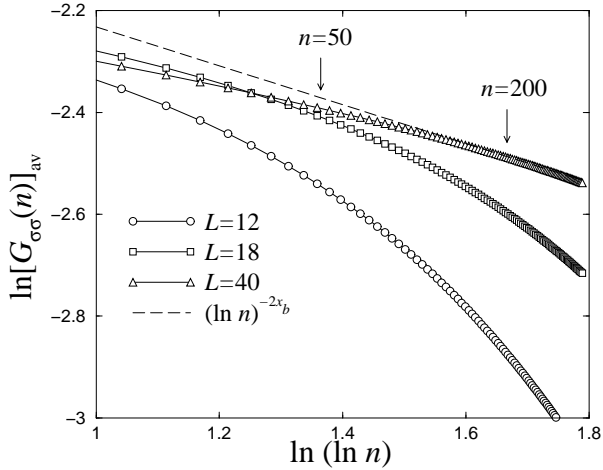,height=6.5cm}}
\vskip 0.2truecm
        \caption{Dynamical spin-spin correlation functions for the $q=3$ Potts
        chain at
	$r=4$ for several strip widths. The number of samples is 30000 for
	$L=18$ and 26000 for $L=40$.}
        \label{fig8b}\end{figure}

Figure~\ref{fig8b} shows the analogous plot of the time correlation
function for $q=3$ and $r=4$. Again for the largest system size 
investigated ($L = 40$) one notices good agreement with the expected
asymptotic limit shown as a thick dashed line. Differently from
Fig.~\ref{fig8} we notice here a crossing of the curves representing
correlation functions for $L = 18$ and $L = 40$. This is we 
believe an effect of the DMRG approach which does not correctly describe the
short distance correlation functions,  since it does
not reproduce the full transfer matrix spectrum. In the other hand, 
when the number of transfer
matrix products becomes large in Eq.~(\ref{classicalcorrelation})
the correlation function is determined by the extremal part of the
spectrum of $T$, which is believed to be well-reproduced by DMRG
methods. Therefore we believe that the crossing of Fig.~\ref{fig8b}
is due to the limits of the DMRG approximation, while results at
larger $n$ should be more accurate.

In conclusion the analysis of the time correlation function
for $q=2$ and $q=3$ also indicates that the universal dynamical properties 
of the disordered quantum Potts chain are independent of $q$.

\section{Conclusions}
\label{sec:con}

We have provided numerical evidences supporting the conclusions of 
Senthil and Majumdar~\cite{senthil96} concerning the existence of a 
unique universality class for all finite value of the $q$-state Potts 
model perturbed by a correlated disorder. For this purpose, we have 
measured the bulk and surface critical exponents and found values 
fully compatible with those exactly calculated by Fisher for the 
RTIC. We point out that a careful numerical study of the critical properties
of disordered quantum
chains requires an accurate average over randomness, 
since the most important source of
 error is due to the disorder average. This effect becomes essential
in some circumstances, possibly leading to wrong exponents, and may even 
dominate other contributions 
like finite-size effects. For that reason, we generated {\it all} the disorder
realizations for chains of relatively small lengths, using the fact that the
finite-size corrections turned out to be quite small (even compared to their
importance in the
corresponding pure models). 
Furthermore, we have checked the expression of the probability 
distribution by an  analysis of its moments: Since the disorder 
average was exact up to the accuracy of the DMRG method, this computation was
allowed even for large moments. The success of this approach is largely due
the occurrence of an optimal moment for which the finite-size corrections appear
to be extremely weak. The method looks promising for further applications to
disordered systems.
 Finally, we have also 
compared the expression of the dynamical correlation function with 
that of the 
RTIC.

Another interesting feature is the illustration, after the work 
of Aizenman and Wehr,\cite{aizenmanwehr89} that correlated disorder
induces a second-order phase transition in the $q-$state Potts model
 with $q>4$. The universality class is however
very robust, in contradistinction with the 
homogeneous two-dimensional disordered
fixed point where the magnetic exponents continuously vary with $q$.

\acknowledgments
We would like to thank R. Couturier for help with the code 
parallelization and D. Karevski for critical reading of the
manuscript. The computations  were performed on the 
{\it SP2} at the CNUSC in Montpellier under project No. C990018, 
and the {\it Power Challenge Array} at the CCH in Nancy. E.C. acknowledges grant
from the Minist\`ere des Affaires Etrang\`eres No 224679A.


\newcommand{\Name}[1]{\rm  #1,}
\newcommand{\And}{\ and\ }
\newcommand{\Review}[1]{\it  #1\rm}
\newcommand{\Vol}[1]{\bf  #1\rm,}
\newcommand{\Year}[1]{\rm  (#1)}
\newcommand{\Page}[1]{\rm  #1}
\newcommand{\Book}[1]{\it  #1\rm}

\def\jpa{J. Phys. A: Math. Gen.}

\def\paper#1#2#3#4#5{#1, #3 {\bf #4}, \rm #5 (#2).}

\end{multicols}

\begin{references}
\bibitem{rq} We followed a different approach than in 
	Ref.~\onlinecite{juozapavicius97}, where
	a bulk magnetic field was added
	to induce a finite magnetization.
\bibitem{rq2} With another definition, namely $[(m_L)^\omega]_{\rm av}\sim
	L^{-\omega\chi(\omega)}$, one would expect in case of
	self-averaging a unique exponent 
	$\chi(\omega)$ for
	all values of $\omega$, i.e. all the moments would scale like the
	average.
\end{references}
\end{document}